\documentclass[aps,prb,twocolumn,superscriptaddress]{revtex4-1}
\usepackage{amsmath,amssymb,graphicx,psfrag}
\usepackage{color,epsfig,rotate}
\usepackage{natbib}

\begin{document}

\preprint{}

\newcommand{\ccs}{CeCu$_{2}$Si$_{2}$}

\title{Observation of two critical points linked to the high field phase B in \ccs}
%


\author{Franziska Weickert}
\email[]{weickert@lanl.gov}
\affiliation{NHMFL, Florida State University, Tallahassee, FL 32310, USA}
\affiliation{MPI for Chemical Physics of Solids, 01187 Dresden, Germany}
\author{Philipp Gegenwart}
\affiliation{EP VI, Center for Electronic Correlations and Magnetism, University of Augsburg, 86159 Augsburg, Germany}
\affiliation{MPI for Chemical Physics of Solids, 01187 Dresden, Germany}
\author{Christoph Geibel}
\affiliation{MPI for Chemical Physics of Solids, 01187 Dresden, Germany}
\author{Wolf Assmus}
\affiliation{Physikalisches Institut, J. W. Goethe-University, 60438 Frankfurt/Main, Germany}
\author{Frank Steglich}
\affiliation{MPI for Chemical Physics of Solids, 01187 Dresden, Germany}
\affiliation{Center of Correlated Matter, Zhejiang University, Hangzhou, China}



\date{\today}

\begin{abstract}
We present thermal expansion and magnetostriction measurements on a \ccs\ single crystal of A/S-type up to 17.9\,T magnetic field applied along the crystallographic $a$-direction ($\Delta L \parallel a \parallel H$) and down to 0.015\,K temperature. We identify clear thermodynamic anomalies at the superconducting transition $T_{c}$ and at two second order transitions $T_{A,B}$ into ordered phases A and B. Our measurements establish for the first time the boundary of phase B at high field and low temperature. No evidence for additional high field phases above B is found up to the maximum field. We speculate based on our experimental results that i) phase B is similar to phase A of spin-density wave type and ii) the first order phase transition between A and B is caused by Fermi surface reconstruction. We furthermore identify a new quantum critical point at $H_{c}\simeq 17$\,T, where $T_{B}$ is suppresssed to zero, and a bicritical point at (0.35\,K, 7.0\,T), where phase lines $T_{A}(H)$ and $T_{B}(H)$ meet.
\end{abstract}

\pacs{}
\keywords{heavy fermion system, unconventional superconductivity, \ccs, thermal expansion, magnetostriction, phase diagram}

\maketitle


\section{\label{intro}Introduction}

\ccs\ is one of the most intriguing heavy fermion (HF) superconductors since its discovery in 1979 \cite{steglich_79}. Since decades, it is strongly believed that superconductivity and magnetism in \ccs\ do not only coexist side by side, but that superconductivity is actually caused by magnetic fluctuations associated with a nearby quantum critical point \cite{kitaoka_87,uemura_89,gegenwart_98}. Strong evidence for this scenario was found by \textit{Stockert et al.} in inelastic neutron scattering (INS) experiments \cite{stockert_11}, where gapped spin excitations inside the superconducting (SC) state were observed. Usually, it is assumed that because of strong onside Coulomb repulsion between $f$ electrons, magnetic mediated superconductivity in heavy fermion systems should result in a SC order parameter with either $d$-wave or $p$-wave symmetry. While this simple approach has been supported by results on CeCoIn$_{5}$ and CeIrIn$_{5}$ systems \cite{movshovich_02}, recent experiments probing the symmetry of the superconducting order parameter in \ccs , such as thermal conductivity, specific heat and magnetization partially conducted under rotational fields find no evidence for nodes in the energy gap \cite{yamashita_17,kittaka_16,takenaka_17}. A recent theoretical study favors a node-less $s^{\pm}$-wave function while taking into account intra as well as strong inter-band magnetic quantum critical scattering \cite{li_18}. On the other hand, \textit{Pang et al.} propose an effective 2-band $d$-wave model \cite{pang_18}, which also explains fully gapped behavior at very low temperatures and is in accordance with a sign change of the SC order parameter as found in INS experiments \cite{stockert_11}. Needless to say, the discussion is ongoing and the relationship between superconductivity and magnetism stays a highly topical area of research, 39 years after the discovery of superconductivity in magnetic \ccs . 

The ground state of \textit{homogeneous} \ccs\ is sensitive to the precise stoichiometry of the sample, because minimal Cu excess or deficiency changes the hybridization between $f$-electrons and conduction electrons \cite{gegenwart_98}. Small deviations from the 1:2:2 ratio produce either an antiferromagnetically ordered A-type (A) ground state, an only superconducting (S) ground state or an A phase that undergoes a transition into superconductivity at lower temperatures (A/S) as depicted schematically in the inset of Fig.\ref{ThermExp}\cite{gegenwart_98}. A/S-type single crystals are closest to the nominal 1:2:2 ratio\cite{steglich_12}. Common to all three types of single crystals is the occurrence of a second field induced phase B.
The phase transition into phase B has been found so far in measurements of the elastic constants\cite{bruls_94}, the resistivity\cite{steglich_00}, and in magnetization experiments\cite{tayama_03}. \textit{Lang et al.} detected a clear anomaly at the onset of phase B in magnetostriction measurements in magnetic fields to 8\,T and temperatures down to 0.25\,K\cite{gonis_99}, which promotes dilatometric measurements as suitable probe to precisely track the phase boundary in even higher magnetic fields.

In the following, we map out the $T-H$ phase diagram of an A/S-type single crystal \ccs\  with magnetic fields applied along the crystallographic $a$-direction ($\Delta L \parallel a \parallel H$). Our work aims at gaining a better understanding of the field-induced phase B, which is currently widely unknown. We furthermore want to explore, if additional phases emerge in higher magnetic fields.
Note, the here presented phase diagram is already mentioned in two review articles \cite{thalmeier_05,zwicknagl_16}, but without showing the actual experimental data.

\section{\label{exp}Methods}
\ccs\ crystalizes in the tetragonal ThCr$_{2}$Si$_{2}$-structure with space group I4/mmm. Large single crystals were grown by crucible free cold boat technique \cite{sun_90}. We observe antiferromagnetic (AFM) order into the A-phase at $T_{A}$=\,0.7\,K and the onset of superconductivity at $T_{c}$= 0.51\,K, which places the A/S sample right at the spot in the phase diagram, where superconductivity and magnetism compete\cite{gegenwart_98,stockert_06b}.

Magnetostriction and thermal expansion of the $L_{0}$(0\,T,\,300\,K)=\,2.26\,mm long sample is measured inside the vacuum chamber of a dilution refrigerator with a capacitive dilatometer. The dilatometer is manufactured from high resistive CuBe alloy to avoid heating effects due to eddy currents during increasing and decreasing magnetic field sweeps. The sample itself was thermally decoupled from the dilatometer with a graphite disk and anchored directly to the mixing chamber with a braid made out of individual silver wires.
This design allows to bypass cooling difficulties caused by a large nuclear Schottky contribution to the specific heat of copper in high magnetic fields.

\section{\label{res}Results}

\begin{figure}
	\includegraphics[width=3.5in]{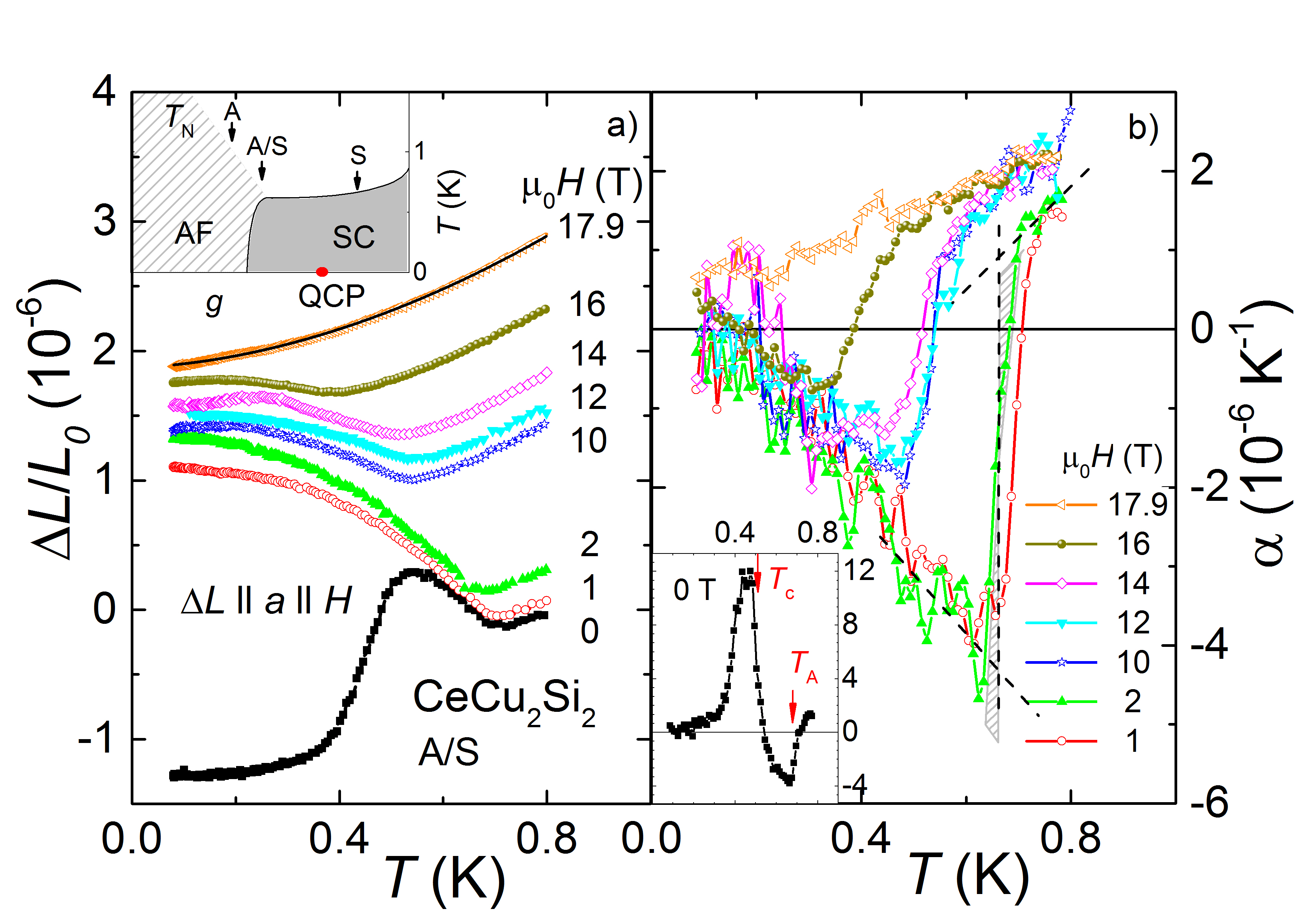}
	\caption{\label{ThermExp} a) Length change $\Delta L/L_{0}$ normalized to the initial value $L_{0}$ versus temperature $T$ from 0.08\,K to 0.8\,K for $\Delta L \parallel a \parallel H$ between 0 and 17.9\,T in \ccs  . The zero field measurement shows two phase transitions whereas measurements in magnetic field only reveal one transition into either phase A or B. The solid line represents a fit to the data at 17.9\,T as expected for a QCP of AFM SDW type in 3D \cite{zhu_03} (for details see Discussion). The corresponding thermal expansion coefficient $\alpha(T)$ for all measurements in magnetic field is shown in b). We estimate the precise transition temperatures $T_{A,B}$ by equal area construction in $\alpha(T)$ as illustrated for the 2\,T data with broken lines and shaded areas. The inset in a) shows the schematic phase diagram $T$ versus $g$. The hybridization $g$ defines the ground state of \ccs\ as A, A/S or S-type as indicated by arrows. The inset in b) displays $\alpha(T)$ in zero field with anomalies at the SC transition $T_{c}=$\,0.51\,K and at the onset of AFM order at $T_{A}=0.7$\,K.}
\end{figure}

At first, we concentrate on thermal expansion measurements between 0.08\,K and 0.8\,K in zero and constant magnetic fields as shown in Fig.\,\ref{ThermExp}. The sample length $\frac{\Delta L}{L_{0}}(T)$ measured in 0\,T expands with increasing temperature inside the SC phase, shows a step-like anomaly at the entrance to phase A with negative slope inside A and a minimum at the transition into the paramagnetic (PM) state followed by an increase of the sample length for $T > 0.7\,$K. Experimental data taken in magnetic fields up to 16\,T show as well a negative $\frac{\Delta L}{L_{0}}(T)$ behavior inside phases A and B and positive length changes when the order is thermally suppressed. We want to emphasize that phases A and B can not be distinguished by the temperature dependence of the thermal expansion. A closer look at the data collected at 14\,T reveals a broad maximum at around 0.25\,K inside phase B. This additional anomaly only occurs in one of the measurements and further investigations are necessary to clarify its origin. The sample length increases monotonically with no obvious anomalies in highest field of 17.9\,T. 

The thermal expansion coefficient is defined as $\alpha(T)= L_{0}^{-1} \times \frac{\partial \Delta L_{a} (T)}{\partial T}$ along one crystallographic direction $a \parallel \Delta L \parallel H$. It measures directly the uniaxial pressure $p_{i}$ dependence of the entropy $S$ via the Maxwell relation $\frac{\partial S}{\partial p_{i}} = -\frac{\partial L_{i}}{\partial T}$ and is therefore well suited to investigate phase transitions and related phenomena with enhanced entropy contributions. Fig.\,\ref{ThermExp}b) shows the thermal expansion coefficient $\alpha(T)$ for the corresponding measurements in a). The transition from paramagnetism into phase A or B is of second order and characterized by a step in the thermal expansion coefficient. We use an equal area construction to get precise values of $T_{A,B}$ as demonstrated for the 2\,T data in Fig.\,\ref{ThermExp}b) and indicated by shaded areas. The inset in Fig.\,\ref{ThermExp}b) shows $\alpha(T)$ at zero field with a first order like anomaly at 0.51\,K that marks the onset of superconductivity.
 
\begin{figure}
	\includegraphics[width=3.5in]{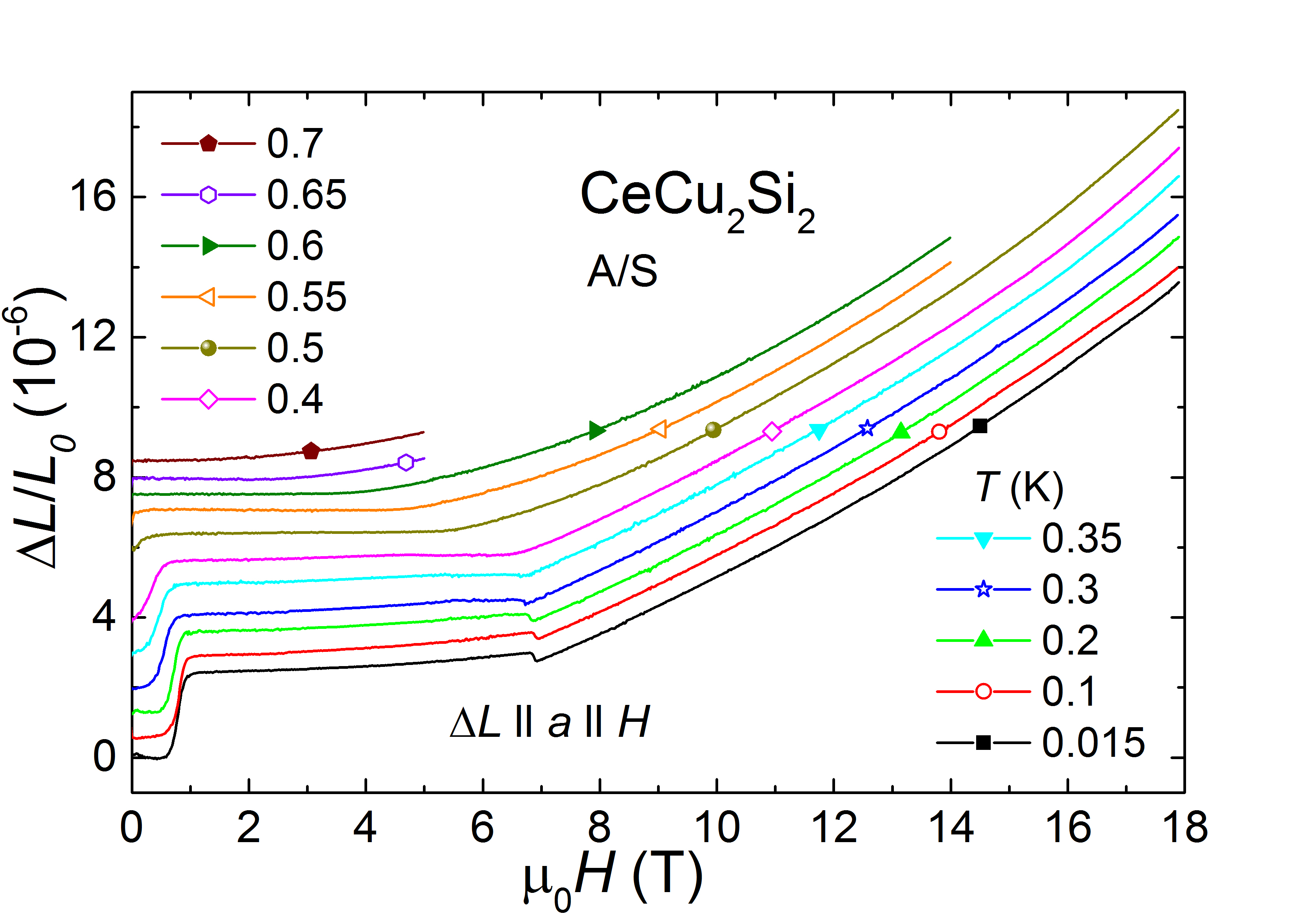}
	\caption{\label{MS_length}Length change $\Delta L/ L_{0}$ normalized to $L_{0}$ versus applied magnetic field $H \parallel \Delta L \parallel a$ between zero and 17.9\,T is shown for temperatures in the range 0.015\,K to 0.7\,K. We observe clear steps at the SC transition at 1\,T and around 7\,T separating phase A from B for lowest temperatures. The anomaly at 1\,T shifts to lower fields with increasing $T$, whereas the step between phase A and B is clearly visible to 0.3\,K and gets less pronounced at higher temperature.}
\end{figure}

The change of sample length $\frac{\Delta L}{L_{0}}(H)$ as a function of magnetic field $H$ for temperatures $T\leq $\,0.7\,K is presented in Fig.\ref{MS_length}). We observe almost no sample expansion for 1\,T $ <\mu_{0}H <$ 7\,T and small quadratic field dependence above 7\,T at lowest temperature. A positive step at 1\,T indicates the suppression of superconductivity and a negative step at $\sim\,7$\,T marks the phase boundary between phases A and B.  

\begin{figure}
	\includegraphics[width=3.5in]{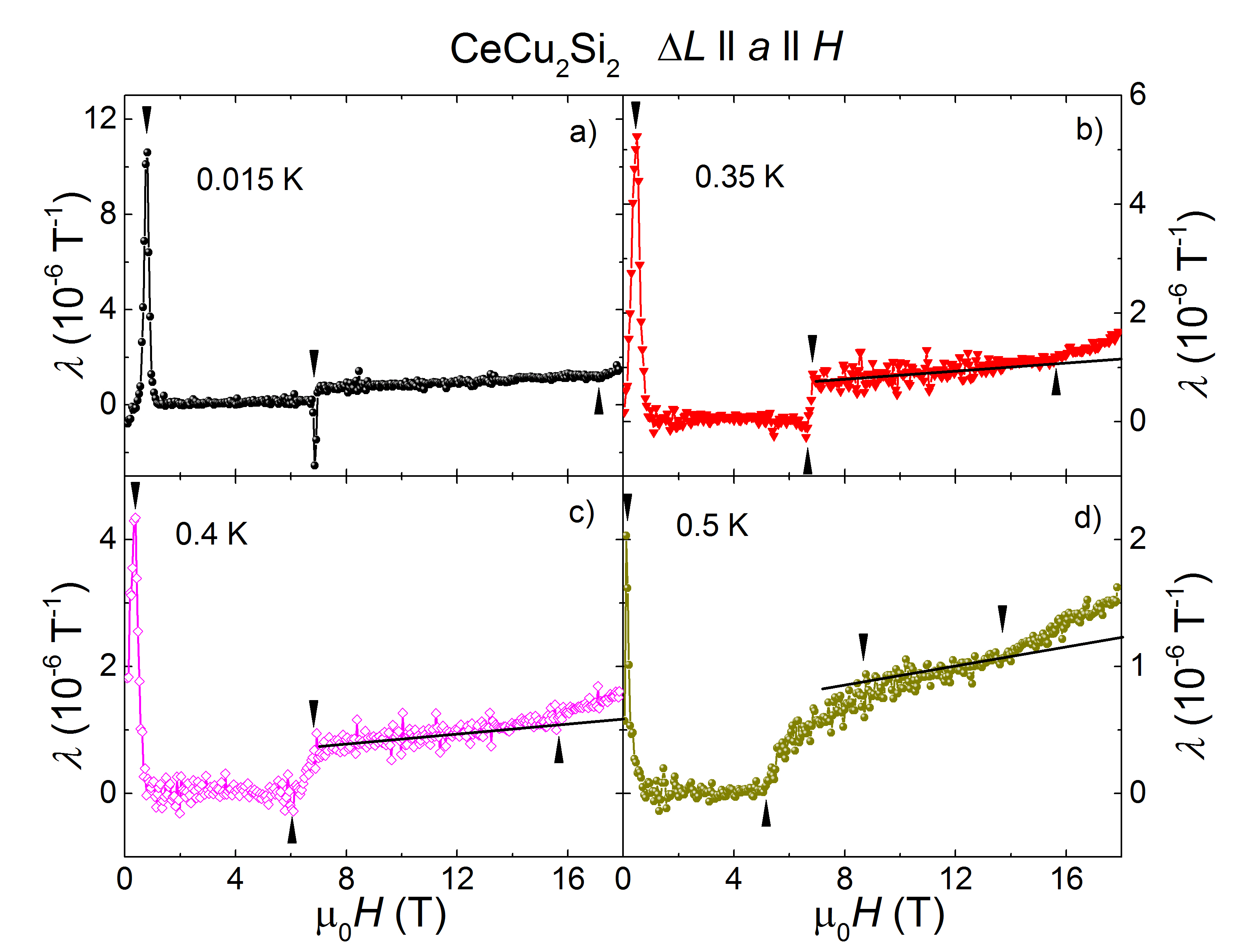}
	\caption{\label{lambda} The magnetostriction coefficient $\lambda(H)$ is shown for 4 selected temperatures a) 0.015\,K, b) 0.35\,K, c) 0.4\,K and d) 0.5\,K illustrating how signatures at the phase transitions (marked with arrows) change with increasing temperature. Please see text for details.}
\end{figure}

The magnetostriction coefficient $\lambda(H)= (L_{0}\mu_{0})^{-1} \times \frac{\partial \Delta L_{a} (T)}{\partial H}$  is the first derivative of the sample length in respect to field.
Panels a) to d) in Fig.\,\ref{lambda} show the evolution of shape and position of the 3 anomalies in $\lambda(H)$ at 1\,T (SC to phase A), 7\,T (phase A to B), and 17\,T (suppression phase B) upon increasing temperature in detail. The sharp positive delta peak in a) that indicates a first order transition between SC ground state and phase A becomes broader and moves to zero field below 0.5\,K. The phase boundary between A and B phase is a sharp negative delta anomaly at lowest temperature suggesting a first order type too. It develops into two separated kinks at 0.35\,K (panel b) that get more distant with increasing temperature as seen in panels c) and d). The suppression of phase B can be inferred from a change of slope in $\lambda(H)$ for fields close to 17\,T (panel a) with slightly reduced critical fields in higher temperatures.  

\section{\label{discussion}Discussion}

\begin{figure}
	\includegraphics[width=3.5in]{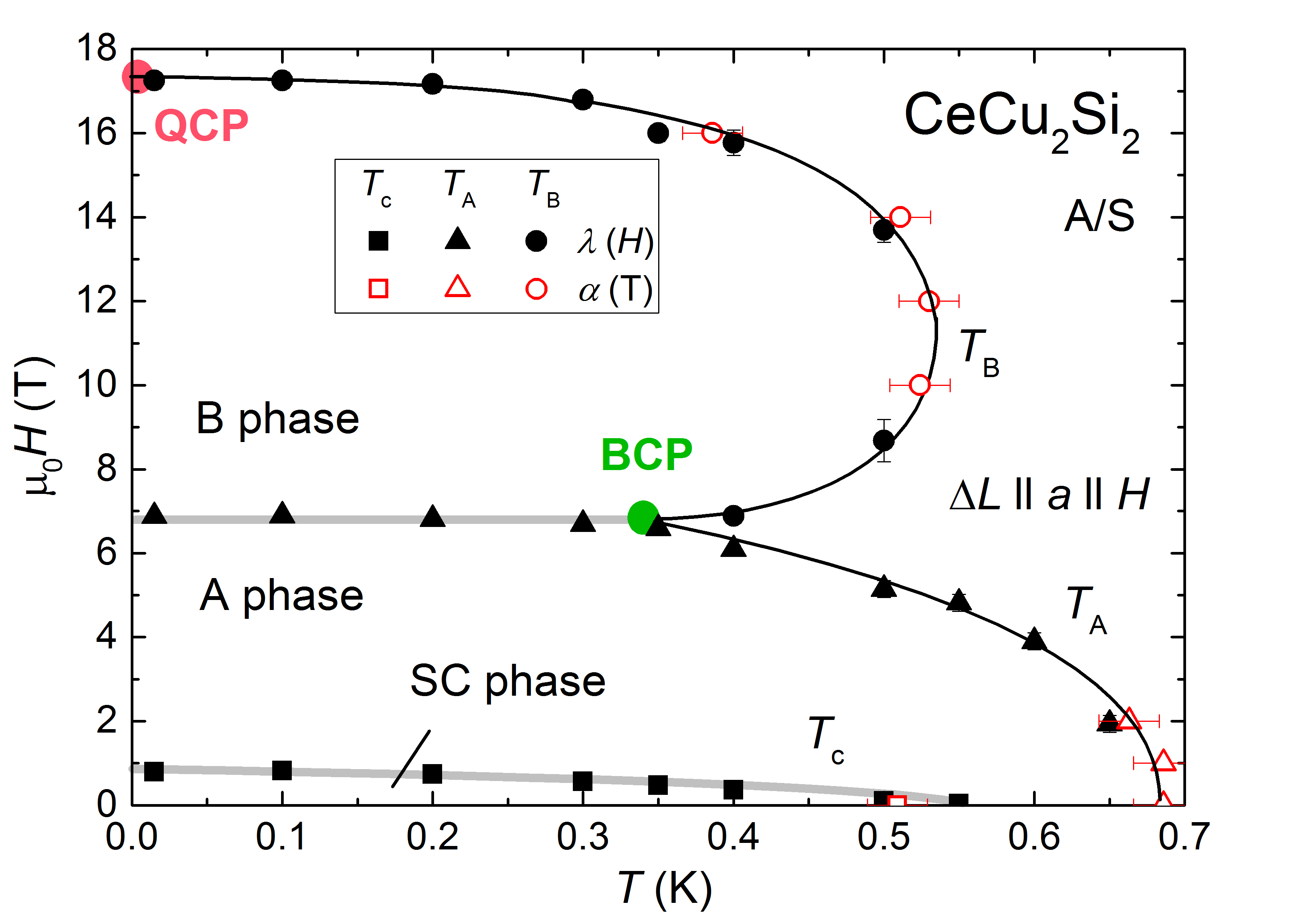}
	\caption{\label{phasedia}$T - H$ phase diagram of an A/S single crystal \ccs\  for $\Delta L \parallel a \parallel H$ as estimated by thermal expansion (open symbols) and magnetostriction (solid symbols) experiments. First order discontinuities are marked with a grey phase boundary and second order phase lines are black, giving rise to a BCP at about (0.35\,K; 7.0\,T) and a QCP at (0\,K; 17\,T).}
\end{figure}

Fig.\,\ref{phasedia} summarizes the results of our thermal expansion and magnetostriction measurements up to 18\,T for magnetic fields applied parallel $a$ in A/S-type \ccs . We find superconductivity below 0.51\,K in zero field, which is suppressed by a magnetic field in excess of 1\,T. The SC ordered phase is surrounded by phase A that evolves into phase B at about 7\,T at lowest temperatures. Phase B is suppressed in fields of 17\,T and higher. A careful inspection of the data did not reveal any hint for additional phases above B, at least up to 17.9\,T. The SC transition and the transition between phase A and B are of first order, marked as grey lines in Fig.\,\ref{phasedia}. All other transitions are of second order (black lines). The phase diagram exhibits a bicritical point (BCP) at about (0.35\,K, 7\,T), where two second order and one first order phase boundaries converge.

In the following, we discuss the nature of phase B. To our knowledge, no experimental data are published probing the local microscopic environment above 7\,T.
Neverteless, certain conclusions on the order in the B-phase can be drawn from macroscopic quantities and similar temperature dependence as found in phase A. Neutron diffraction experiments carried out inside phase A verify incommensurate antiferromagnetic (AFM) order of spin density wave (SDW) type with (0.215, 0.215, 0.530) ordering vector, which is caused by renormalized Fermi-surface (FS) nesting\cite{stockert_04} of the heavy bands. The ordered moment is about $0.1\,\mu_{B}$. $\mu$SR experiments in zero field suggest phase separation of AFM and SC regions at the transition $T_{c}$ with increasing SC volume for decreasing temperatures\cite{stockert_06b}. 

The anomalies in $\alpha(T)$ are very similar at the phase boundaries $T_{A}(H)$ and $T_{B}(H)$ and of second-order type. This suggests a change of symmetry between PM state and both phases A and B, respectively. It is microscopically proven for phase A\cite{stockert_04}. Thermal expansion inside phase B has a similar temperature dependence (negative $\alpha(T)$) as in A, compare Fig.\,\ref{ThermExp}. The transition between phase A and B is most likely a first order discontinuity as indicated by the step in the magnetostriction. Unfortunately, we have only measured at a certain temperature in one field direction, either increasing field or decreasing field. Therefore, we can not identify possible hysteretic behavior, which is often observed at first order transitions. The magnetostriction coefficient $\lambda (H)$ is almost zero in phase A and therefore slightly different from the small linear field dependence as found in the B-phase. It is worth to note that no universal field dependence of $\lambda(H)$ for AFM ordering is expected. It rather depends on the shape and anisotropy of the energy dispersion function for a specific type of magnetic order.    

Now, we will have a look at other thermodynamic quantities. Magnetization experiments up to 11.5\,T ($H \parallel a$) carried out by \textit{Tayama et al.}\cite{tayama_03} reveal a linear increase of the magnetization $M(H)$ inside phase A as well as inside B with similar slope. A step is observed  in $M(H)$ at the transition between both phases at lowest temperatures, which is also an indication for its first order nature. There is only a tiny increase of the magnetization by $2 \times 10^{-3}\mu_{B}$ on going from phase A to B. This can be caused either by a small change of the magnetic moment or by a large moment change that is predominantly screened by the AFM order. The later case seems to be less likely, because large ordered moments were not observed in neutron scattering experiments inside phase B \cite{stockert_18}.  

The results in thermal expansion, magnetostriction as well as magnetization experiments lead us to the conclusion that phase B is similar to phase A, i.e., being of SDW AFM order with a small ordered moment in the order of $\sim 0.1 \mu_{B}$.

Renomalized band structure calculations in zero field \cite{zwicknagl_93} estimate separate sheets of Fermi surface for light and heavy quasiparticles. Relatively light quasiparticles exhibiting 5 times the bare electron mass $m_{e}$ are verified in dHvA experiments.\cite{hunt_90} Heavy quasiparticles ($500 m_{e}$) are expected on a quasi 2-dimensional Fermi surface in the shape of warped cylinders along the $c$-axis and small pockets. 
\textit{Zwicknagl et al.} \cite{zwicknagl_93,zwicknagl_16} suggest that the transition between phase A and B is a Lifshitz transition and that the FS topology changes from quasi 2D to 3D. Unfortunately, it is impossible to test the change of dimensionality by quantum oscillation measurements, because of the heavy mass of the quasi-particles. \textit{Hunt et al.} reported only a small change in the dHvA-frequency and mass of the light bands from 171\,T, 4.62 $m_{e}$ inside the A-phase to 162\,T, 5.15 $m_{e}$ inside phase B.   \cite{hunt_90} We wish to emphasize that our observation of a first order phase transition between phases A and B would be consistent with a Lifshitz transition\cite{lifshitz_60}.

A possible alternative scenario to a Lifshitz transition occurring at 7\,T is domain reorientation. \textit{Stockert et al.} find indeed in neutron diffraction experiments at $H = 0$ symmetry equivalent peak positions that point to the existence of AFM domains inside phase A.\cite{stockert_04} Assuming the transition at 7\,T do be solely due to domain reorientation, clear magnetic Bragg peaks are expected inside phase B at or close to some of the Bragg peaks observed in phase A. However elastic neutron scattering experiments in phase B didn´t reveal any magnetic Bragg peaks.\cite{stockert_18} This excludes the transition at 7\,T to be merely due to domain reorientation.

Finally, we comment on the second order phase line $T_{B}(H)$ that is suppressed in magnetic fields and gives rise to a QCP with critical field $H_{c} \cong  17$\,T, see Fig.\,\ref{phasedia}. The thermal expansion measurement that comes closest to $H_{c}$ is the one carried out in 17.9\,T. Fig.\,\ref{ThermExp}a) shows a power law fit $\Delta L/L = a_{0} + a_{1} T^{3/2} + a_{2} T^{2}$ (solid line) with reasonable agreement within scattering of the experimental data and $a_{1}, a_{2} > 0$. The $T^{3/2}$-term is hereby expected for an AFM SDW QCP in 3 dimensions \cite{zhu_03} and the quadratic term typical for (non-critical) Landau-Fermi liquid contributions. While this observation is only a first hint for quantum critical behavior occurring close to this newly discovered magnetic field-induced QCP, it is likely to stimulate further experimental investigations of other bulk as well as microscopic properties.

Note, the combination of thermal expansion measurements together with specific heat data $C(T)$ is an extremely successful method to detect and classify quantum critical behavior, because the Gr\"uneisen ratio $\Gamma \sim \alpha(T)/ C(T)$, a measure of the relevant energy scale, diverges at QCPs with certain power laws \cite{zhu_03}. This approach has been applied i.e. to classify the zero pressure QCP in CeNi$_{2}$Ge$_{2}$ as 3-dimensional QCP of AFM-SDW type \cite{kuechler_03} with same $\Delta L/L \sim T^{3/2}$ critical contribution to the length change as observed in this new field-induced QCP in \ccs . In contrast, the SDW description was excluded based on this analysis for other heavy-fermion systems such as	CeCu$_{5.8}$Ag$_{0.2}$ \cite{kuechler_04a} or YbRh$_{2}$Si$_{2}$. \cite{kuechler_03,kuechler_04b}






\section{\label{sum}Summary}

In conclusion, we have used thermal expansion and magnetostriction experiments to establish the complete $H-T$-phase diagram of A/S-type \ccs\ up to 17.9\,T. We confirm the existence of three different types of ordering; superconductivity, and magnetic phase A and B, with the SC phase occurring below 0.51\,K and in magnetic fields up to 1\,T. Phase A is stable up to 7\,T showing a weak first order transition into phase B. Our dilatometric measurements support the picture that phase B is of similar SDW-type as phase A. We identify two new field-induced critical points in the phase diagram. A QCP is observed at $H_{c} \cong 17$\,T with hints of quantum critical behavior in the thermal expansion for $H\ge H_{c}$. We furthermore identify a bicritical point at finite temperature at 0.35\,K and 7\,T, where two second order phase lines $T_{A}(H)$ and $T_{B}(H)$ and one first order phase line merge.

\begin{acknowledgments}
A portion of this work was performed at the National High Magnetic Field Laboratory, which is supported by National Science Foundation Cooperative Agreement No. DMR-1157490,  the State of Florida and the United States Department of Energy.
\end{acknowledgments}

\bibliography{article_20180425}

\begin{thebibliography}{29}%
\makeatletter
\providecommand \@ifxundefined [1]{%
 \@ifx{#1\undefined}
}%
\providecommand \@ifnum [1]{%
 \ifnum #1\expandafter \@firstoftwo
 \else \expandafter \@secondoftwo
 \fi
}%
\providecommand \@ifx [1]{%
 \ifx #1\expandafter \@firstoftwo
 \else \expandafter \@secondoftwo
 \fi
}%
\providecommand \natexlab [1]{#1}%
\providecommand \enquote  [1]{``#1''}%
\providecommand \bibnamefont  [1]{#1}%
\providecommand \bibfnamefont [1]{#1}%
\providecommand \citenamefont [1]{#1}%
\providecommand \href@noop [0]{\@secondoftwo}%
\providecommand \href [0]{\begingroup \@sanitize@url \@href}%
\providecommand \@href[1]{\@@startlink{#1}\@@href}%
\providecommand \@@href[1]{\endgroup#1\@@endlink}%
\providecommand \@sanitize@url [0]{\catcode `\\12\catcode `\$12\catcode
  `\&12\catcode `\#12\catcode `\^12\catcode `\_12\catcode `\%12\relax}%
\providecommand \@@startlink[1]{}%
\providecommand \@@endlink[0]{}%
\providecommand \url  [0]{\begingroup\@sanitize@url \@url }%
\providecommand \@url [1]{\endgroup\@href {#1}{\urlprefix }}%
\providecommand \urlprefix  [0]{URL }%
\providecommand \Eprint [0]{\href }%
\providecommand \doibase [0]{http://dx.doi.org/}%
\providecommand \selectlanguage [0]{\@gobble}%
\providecommand \bibinfo  [0]{\@secondoftwo}%
\providecommand \bibfield  [0]{\@secondoftwo}%
\providecommand \translation [1]{[#1]}%
\providecommand \BibitemOpen [0]{}%
\providecommand \bibitemStop [0]{}%
\providecommand \bibitemNoStop [0]{.\EOS\space}%
\providecommand \EOS [0]{\spacefactor3000\relax}%
\providecommand \BibitemShut  [1]{\csname bibitem#1\endcsname}%
\let\auto@bib@innerbib\@empty
\bibitem [{\citenamefont {Steglich}\ \emph {et~al.}(1979)\citenamefont
  {Steglich}, \citenamefont {Aarts}, \citenamefont {Bredl}, \citenamefont
  {Lieke}, \citenamefont {Meschede}, \citenamefont {Franz},\ and\ \citenamefont
  {Sch\"afer}}]{steglich_79}%
  \BibitemOpen
  \bibfield  {author} {\bibinfo {author} {\bibfnamefont {F.}~\bibnamefont
  {Steglich}}, \bibinfo {author} {\bibfnamefont {J.}~\bibnamefont {Aarts}},
  \bibinfo {author} {\bibfnamefont {C.~D.}\ \bibnamefont {Bredl}}, \bibinfo
  {author} {\bibfnamefont {W.}~\bibnamefont {Lieke}}, \bibinfo {author}
  {\bibfnamefont {D.}~\bibnamefont {Meschede}}, \bibinfo {author}
  {\bibfnamefont {W.}~\bibnamefont {Franz}}, \ and\ \bibinfo {author}
  {\bibfnamefont {H.}~\bibnamefont {Sch\"afer}},\ }\href@noop {} {\bibfield
  {journal} {\bibinfo  {journal} {Physical Review Letters}\ }\textbf {\bibinfo
  {volume} {43}},\ \bibinfo {pages} {1892} (\bibinfo {year}
  {1979})}\BibitemShut {NoStop}%
\bibitem [{\citenamefont {Kitaoka}\ \emph {et~al.}(1987)\citenamefont
  {Kitaoka}, \citenamefont {Yamada}, \citenamefont {ich Ueda}, \citenamefont
  {Kohori}, \citenamefont {Kohara}, \citenamefont {Oda},\ and\ \citenamefont
  {Asayama}}]{kitaoka_87}%
  \BibitemOpen
  \bibfield  {author} {\bibinfo {author} {\bibfnamefont {Y.}~\bibnamefont
  {Kitaoka}}, \bibinfo {author} {\bibfnamefont {H.}~\bibnamefont {Yamada}},
  \bibinfo {author} {\bibfnamefont {K.}~\bibnamefont {ich Ueda}}, \bibinfo
  {author} {\bibfnamefont {Y.}~\bibnamefont {Kohori}}, \bibinfo {author}
  {\bibfnamefont {T.}~\bibnamefont {Kohara}}, \bibinfo {author} {\bibfnamefont
  {Y.}~\bibnamefont {Oda}}, \ and\ \bibinfo {author} {\bibfnamefont
  {K.}~\bibnamefont {Asayama}},\ }\href
  {http://iopscience.iop.org/1347-4065/26/S3-2/1221} {\bibfield  {journal}
  {\bibinfo  {journal} {Japanese Journal of Applied Physics}\ }\textbf
  {\bibinfo {volume} {26}},\ \bibinfo {pages} {1221} (\bibinfo {year}
  {1987})}\BibitemShut {NoStop}%
\bibitem [{\citenamefont {Uemura}\ \emph {et~al.}(1989)\citenamefont {Uemura},
  \citenamefont {Kossler}, \citenamefont {Yu}, \citenamefont {Schone},
  \citenamefont {Kempton}, \citenamefont {Stronach}, \citenamefont {Barth},
  \citenamefont {Gygax}, \citenamefont {Hitti}, \citenamefont {Schenck},
  \citenamefont {Baines}, \citenamefont {Lankford}, \citenamefont
  {\ifmmode~\bar{O}\else \={O}\fi{}nuki},\ and\ \citenamefont
  {Komatsubara}}]{uemura_89}%
  \BibitemOpen
  \bibfield  {author} {\bibinfo {author} {\bibfnamefont {Y.~J.}\ \bibnamefont
  {Uemura}}, \bibinfo {author} {\bibfnamefont {W.~J.}\ \bibnamefont {Kossler}},
  \bibinfo {author} {\bibfnamefont {X.~H.}\ \bibnamefont {Yu}}, \bibinfo
  {author} {\bibfnamefont {H.~E.}\ \bibnamefont {Schone}}, \bibinfo {author}
  {\bibfnamefont {J.~R.}\ \bibnamefont {Kempton}}, \bibinfo {author}
  {\bibfnamefont {C.~E.}\ \bibnamefont {Stronach}}, \bibinfo {author}
  {\bibfnamefont {S.}~\bibnamefont {Barth}}, \bibinfo {author} {\bibfnamefont
  {F.~N.}\ \bibnamefont {Gygax}}, \bibinfo {author} {\bibfnamefont
  {B.}~\bibnamefont {Hitti}}, \bibinfo {author} {\bibfnamefont
  {A.}~\bibnamefont {Schenck}}, \bibinfo {author} {\bibfnamefont
  {C.}~\bibnamefont {Baines}}, \bibinfo {author} {\bibfnamefont {W.~F.}\
  \bibnamefont {Lankford}}, \bibinfo {author} {\bibfnamefont {Y.}~\bibnamefont
  {\ifmmode~\bar{O}\else \={O}\fi{}nuki}}, \ and\ \bibinfo {author}
  {\bibfnamefont {T.}~\bibnamefont {Komatsubara}},\ }\href {\doibase
  10.1103/PhysRevB.39.4726} {\bibfield  {journal} {\bibinfo  {journal}
  {Physical Review B}\ }\textbf {\bibinfo {volume} {39}},\ \bibinfo {pages}
  {4726} (\bibinfo {year} {1989})}\BibitemShut {NoStop}%
\bibitem [{\citenamefont {Gegenwart}\ \emph {et~al.}(1998)\citenamefont
  {Gegenwart}, \citenamefont {Langhammer}, \citenamefont {Geibel},
  \citenamefont {Helfrich}, \citenamefont {Lang}, \citenamefont {Sparn},
  \citenamefont {Steglich}, \citenamefont {Horn}, \citenamefont {Donnevert},
  \citenamefont {Link},\ and\ \citenamefont {Assmus}}]{gegenwart_98}%
  \BibitemOpen
  \bibfield  {author} {\bibinfo {author} {\bibfnamefont {P.}~\bibnamefont
  {Gegenwart}}, \bibinfo {author} {\bibfnamefont {C.}~\bibnamefont
  {Langhammer}}, \bibinfo {author} {\bibfnamefont {C.}~\bibnamefont {Geibel}},
  \bibinfo {author} {\bibfnamefont {R.}~\bibnamefont {Helfrich}}, \bibinfo
  {author} {\bibfnamefont {M.}~\bibnamefont {Lang}}, \bibinfo {author}
  {\bibfnamefont {G.}~\bibnamefont {Sparn}}, \bibinfo {author} {\bibfnamefont
  {F.}~\bibnamefont {Steglich}}, \bibinfo {author} {\bibfnamefont
  {R.}~\bibnamefont {Horn}}, \bibinfo {author} {\bibfnamefont {L.}~\bibnamefont
  {Donnevert}}, \bibinfo {author} {\bibfnamefont {A.}~\bibnamefont {Link}}, \
  and\ \bibinfo {author} {\bibfnamefont {W.}~\bibnamefont {Assmus}},\ }\href
  {\doibase 10.1103/PhysRevLett.81.1501} {\bibfield  {journal} {\bibinfo
  {journal} {Physical Review Letters}\ }\textbf {\bibinfo {volume} {81}},\
  \bibinfo {pages} {1501} (\bibinfo {year} {1998})}\BibitemShut {NoStop}%
\bibitem [{\citenamefont {Stockert}\ \emph {et~al.}(2011)\citenamefont
  {Stockert}, \citenamefont {Arndt}, \citenamefont {Faulhaber}, \citenamefont
  {Geibel}, \citenamefont {Jeevan}, \citenamefont {Kirchner}, \citenamefont
  {L\"owenhaupt}, \citenamefont {Schmalzl}, \citenamefont {Schmidt},
  \citenamefont {Si},\ and\ \citenamefont {Steglich}}]{stockert_11}%
  \BibitemOpen
  \bibfield  {author} {\bibinfo {author} {\bibfnamefont {O.}~\bibnamefont
  {Stockert}}, \bibinfo {author} {\bibfnamefont {J.}~\bibnamefont {Arndt}},
  \bibinfo {author} {\bibfnamefont {E.}~\bibnamefont {Faulhaber}}, \bibinfo
  {author} {\bibfnamefont {C.}~\bibnamefont {Geibel}}, \bibinfo {author}
  {\bibfnamefont {H.~S.}\ \bibnamefont {Jeevan}}, \bibinfo {author}
  {\bibfnamefont {S.}~\bibnamefont {Kirchner}}, \bibinfo {author}
  {\bibfnamefont {M.}~\bibnamefont {L\"owenhaupt}}, \bibinfo {author}
  {\bibfnamefont {K.}~\bibnamefont {Schmalzl}}, \bibinfo {author}
  {\bibfnamefont {W.}~\bibnamefont {Schmidt}}, \bibinfo {author} {\bibfnamefont
  {Q.}~\bibnamefont {Si}}, \ and\ \bibinfo {author} {\bibfnamefont
  {F.}~\bibnamefont {Steglich}},\ }\href {\doibase 10.1038/nphys1852}
  {\bibfield  {journal} {\bibinfo  {journal} {Nature Physics}\ }\textbf
  {\bibinfo {volume} {7}},\ \bibinfo {pages} {119} (\bibinfo {year}
  {2011})}\BibitemShut {NoStop}%
\bibitem [{\citenamefont {Movshovich}\ \emph {et~al.}(2002)\citenamefont
  {Movshovich}, \citenamefont {Bianchi}, \citenamefont {Jaime}, \citenamefont
  {Hundley}, \citenamefont {Thompson}, \citenamefont {Curro}, \citenamefont
  {Hammel}, \citenamefont {Fisk}, \citenamefont {Pagliuso},\ and\ \citenamefont
  {Sarrao}}]{movshovich_02}%
  \BibitemOpen
  \bibfield  {author} {\bibinfo {author} {\bibfnamefont {R.}~\bibnamefont
  {Movshovich}}, \bibinfo {author} {\bibfnamefont {A.}~\bibnamefont {Bianchi}},
  \bibinfo {author} {\bibfnamefont {M.}~\bibnamefont {Jaime}}, \bibinfo
  {author} {\bibfnamefont {M.}~\bibnamefont {Hundley}}, \bibinfo {author}
  {\bibfnamefont {J.}~\bibnamefont {Thompson}}, \bibinfo {author}
  {\bibfnamefont {N.}~\bibnamefont {Curro}}, \bibinfo {author} {\bibfnamefont
  {P.}~\bibnamefont {Hammel}}, \bibinfo {author} {\bibfnamefont
  {Z.}~\bibnamefont {Fisk}}, \bibinfo {author} {\bibfnamefont {P.}~\bibnamefont
  {Pagliuso}}, \ and\ \bibinfo {author} {\bibfnamefont {J.}~\bibnamefont
  {Sarrao}},\ }\href {\doibase https://doi.org/10.1016/S0921-4526(01)01062-6}
  {\bibfield  {journal} {\bibinfo  {journal} {Physica B: Condensed Matter}\
  }\textbf {\bibinfo {volume} {312-313}},\ \bibinfo {pages} {7 } (\bibinfo
  {year} {2002})},\ \bibinfo {note} {the International Conference on Strongly
  Correlated Electron Systems}\BibitemShut {NoStop}%
\bibitem [{\citenamefont {Yamashita}\ \emph {et~al.}(2017)\citenamefont
  {Yamashita}, \citenamefont {Takenaka}, \citenamefont {Tokiwa}, \citenamefont
  {Wilcox}, \citenamefont {Mizukami}, \citenamefont {Terazawa}, \citenamefont
  {Kasahara}, \citenamefont {Kittaka}, \citenamefont {Sakakibara},
  \citenamefont {Konczykowski}, \citenamefont {Seiro}, \citenamefont {Jeevan},
  \citenamefont {Geibel}, \citenamefont {Putzke}, \citenamefont {Onishi},
  \citenamefont {Ikeda}, \citenamefont {Carrington}, \citenamefont
  {Shibauchi},\ and\ \citenamefont {Matsuda}}]{yamashita_17}%
  \BibitemOpen
  \bibfield  {author} {\bibinfo {author} {\bibfnamefont {T.}~\bibnamefont
  {Yamashita}}, \bibinfo {author} {\bibfnamefont {T.}~\bibnamefont {Takenaka}},
  \bibinfo {author} {\bibfnamefont {Y.}~\bibnamefont {Tokiwa}}, \bibinfo
  {author} {\bibfnamefont {J.~A.}\ \bibnamefont {Wilcox}}, \bibinfo {author}
  {\bibfnamefont {Y.}~\bibnamefont {Mizukami}}, \bibinfo {author}
  {\bibfnamefont {D.}~\bibnamefont {Terazawa}}, \bibinfo {author}
  {\bibfnamefont {Y.}~\bibnamefont {Kasahara}}, \bibinfo {author}
  {\bibfnamefont {S.}~\bibnamefont {Kittaka}}, \bibinfo {author} {\bibfnamefont
  {T.}~\bibnamefont {Sakakibara}}, \bibinfo {author} {\bibfnamefont
  {M.}~\bibnamefont {Konczykowski}}, \bibinfo {author} {\bibfnamefont
  {S.}~\bibnamefont {Seiro}}, \bibinfo {author} {\bibfnamefont {H.~S.}\
  \bibnamefont {Jeevan}}, \bibinfo {author} {\bibfnamefont {C.}~\bibnamefont
  {Geibel}}, \bibinfo {author} {\bibfnamefont {C.}~\bibnamefont {Putzke}},
  \bibinfo {author} {\bibfnamefont {T.}~\bibnamefont {Onishi}}, \bibinfo
  {author} {\bibfnamefont {H.}~\bibnamefont {Ikeda}}, \bibinfo {author}
  {\bibfnamefont {A.}~\bibnamefont {Carrington}}, \bibinfo {author}
  {\bibfnamefont {T.}~\bibnamefont {Shibauchi}}, \ and\ \bibinfo {author}
  {\bibfnamefont {Y.}~\bibnamefont {Matsuda}},\ }\href {\doibase
  10.1126/sciadv.1601667} {\bibfield  {journal} {\bibinfo  {journal} {Science
  Advances}\ }\textbf {\bibinfo {volume} {3}} (\bibinfo {year} {2017}),\
  10.1126/sciadv.1601667}\BibitemShut {NoStop}%
\bibitem [{\citenamefont {Kittaka}\ \emph {et~al.}(2016)\citenamefont
  {Kittaka}, \citenamefont {Aoki}, \citenamefont {Shimura}, \citenamefont
  {Sakakibara}, \citenamefont {Seiro}, \citenamefont {Geibel}, \citenamefont
  {Steglich}, \citenamefont {Tsutsumi}, \citenamefont {Ikeda},\ and\
  \citenamefont {Machida}}]{kittaka_16}%
  \BibitemOpen
  \bibfield  {author} {\bibinfo {author} {\bibfnamefont {S.}~\bibnamefont
  {Kittaka}}, \bibinfo {author} {\bibfnamefont {Y.}~\bibnamefont {Aoki}},
  \bibinfo {author} {\bibfnamefont {Y.}~\bibnamefont {Shimura}}, \bibinfo
  {author} {\bibfnamefont {T.}~\bibnamefont {Sakakibara}}, \bibinfo {author}
  {\bibfnamefont {S.}~\bibnamefont {Seiro}}, \bibinfo {author} {\bibfnamefont
  {C.}~\bibnamefont {Geibel}}, \bibinfo {author} {\bibfnamefont
  {F.}~\bibnamefont {Steglich}}, \bibinfo {author} {\bibfnamefont
  {Y.}~\bibnamefont {Tsutsumi}}, \bibinfo {author} {\bibfnamefont
  {H.}~\bibnamefont {Ikeda}}, \ and\ \bibinfo {author} {\bibfnamefont
  {K.}~\bibnamefont {Machida}},\ }\href {\doibase 10.1103/PhysRevB.94.054514}
  {\bibfield  {journal} {\bibinfo  {journal} {Physical Revew B}\ }\textbf
  {\bibinfo {volume} {94}},\ \bibinfo {pages} {054514} (\bibinfo {year}
  {2016})}\BibitemShut {NoStop}%
\bibitem [{\citenamefont {Takenaka}\ \emph {et~al.}(2017)\citenamefont
  {Takenaka}, \citenamefont {Mizukami}, \citenamefont {Wilcox}, \citenamefont
  {Konczykowski}, \citenamefont {Seiro}, \citenamefont {Geibel}, \citenamefont
  {Tokiwa}, \citenamefont {Kasahara}, \citenamefont {Putzke}, \citenamefont
  {Matsuda}, \citenamefont {Carrington},\ and\ \citenamefont
  {Shibauchi}}]{takenaka_17}%
  \BibitemOpen
  \bibfield  {author} {\bibinfo {author} {\bibfnamefont {T.}~\bibnamefont
  {Takenaka}}, \bibinfo {author} {\bibfnamefont {Y.}~\bibnamefont {Mizukami}},
  \bibinfo {author} {\bibfnamefont {J.~A.}\ \bibnamefont {Wilcox}}, \bibinfo
  {author} {\bibfnamefont {M.}~\bibnamefont {Konczykowski}}, \bibinfo {author}
  {\bibfnamefont {S.}~\bibnamefont {Seiro}}, \bibinfo {author} {\bibfnamefont
  {C.}~\bibnamefont {Geibel}}, \bibinfo {author} {\bibfnamefont
  {Y.}~\bibnamefont {Tokiwa}}, \bibinfo {author} {\bibfnamefont
  {Y.}~\bibnamefont {Kasahara}}, \bibinfo {author} {\bibfnamefont
  {C.}~\bibnamefont {Putzke}}, \bibinfo {author} {\bibfnamefont
  {Y.}~\bibnamefont {Matsuda}}, \bibinfo {author} {\bibfnamefont
  {A.}~\bibnamefont {Carrington}}, \ and\ \bibinfo {author} {\bibfnamefont
  {T.}~\bibnamefont {Shibauchi}},\ }\href {\doibase
  10.1103/PhysRevLett.119.077001} {\bibfield  {journal} {\bibinfo  {journal}
  {Physical Review Letters}\ }\textbf {\bibinfo {volume} {119}},\ \bibinfo
  {pages} {077001} (\bibinfo {year} {2017})}\BibitemShut {NoStop}%
\bibitem [{\citenamefont {Li}\ \emph {et~al.}(2018)\citenamefont {Li},
  \citenamefont {Liu}, \citenamefont {Fu}, \citenamefont {Chen}, \citenamefont
  {Yang},\ and\ \citenamefont {Yang}}]{li_18}%
  \BibitemOpen
  \bibfield  {author} {\bibinfo {author} {\bibfnamefont {Y.}~\bibnamefont
  {Li}}, \bibinfo {author} {\bibfnamefont {M.}~\bibnamefont {Liu}}, \bibinfo
  {author} {\bibfnamefont {Z.}~\bibnamefont {Fu}}, \bibinfo {author}
  {\bibfnamefont {X.}~\bibnamefont {Chen}}, \bibinfo {author} {\bibfnamefont
  {F.}~\bibnamefont {Yang}}, \ and\ \bibinfo {author} {\bibfnamefont {Y.-f.}\
  \bibnamefont {Yang}},\ }\href {\doibase 10.1103/PhysRevLett.120.217001}
  {\bibfield  {journal} {\bibinfo  {journal} {Phys. Rev. Lett.}\ }\textbf
  {\bibinfo {volume} {120}},\ \bibinfo {pages} {217001} (\bibinfo {year}
  {2018})}\BibitemShut {NoStop}%
\bibitem [{\citenamefont {Pang}\ \emph {et~al.}(2018)\citenamefont {Pang},
  \citenamefont {Smidman}, \citenamefont {Zhang}, \citenamefont {Jiao},
  \citenamefont {Weng}, \citenamefont {Nica}, \citenamefont {Chen},
  \citenamefont {Jiang}, \citenamefont {Zhang}, \citenamefont {Xie},
  \citenamefont {Jeevan}, \citenamefont {Lee}, \citenamefont {Gegenwart},
  \citenamefont {Steglich}, \citenamefont {Si},\ and\ \citenamefont
  {Yuan}}]{pang_18}%
  \BibitemOpen
  \bibfield  {author} {\bibinfo {author} {\bibfnamefont {G.}~\bibnamefont
  {Pang}}, \bibinfo {author} {\bibfnamefont {M.}~\bibnamefont {Smidman}},
  \bibinfo {author} {\bibfnamefont {J.}~\bibnamefont {Zhang}}, \bibinfo
  {author} {\bibfnamefont {L.}~\bibnamefont {Jiao}}, \bibinfo {author}
  {\bibfnamefont {Z.}~\bibnamefont {Weng}}, \bibinfo {author} {\bibfnamefont
  {E.~M.}\ \bibnamefont {Nica}}, \bibinfo {author} {\bibfnamefont
  {Y.}~\bibnamefont {Chen}}, \bibinfo {author} {\bibfnamefont {W.}~\bibnamefont
  {Jiang}}, \bibinfo {author} {\bibfnamefont {Y.}~\bibnamefont {Zhang}},
  \bibinfo {author} {\bibfnamefont {W.}~\bibnamefont {Xie}}, \bibinfo {author}
  {\bibfnamefont {H.~S.}\ \bibnamefont {Jeevan}}, \bibinfo {author}
  {\bibfnamefont {H.}~\bibnamefont {Lee}}, \bibinfo {author} {\bibfnamefont
  {P.}~\bibnamefont {Gegenwart}}, \bibinfo {author} {\bibfnamefont
  {F.}~\bibnamefont {Steglich}}, \bibinfo {author} {\bibfnamefont
  {Q.}~\bibnamefont {Si}}, \ and\ \bibinfo {author} {\bibfnamefont
  {H.}~\bibnamefont {Yuan}},\ }\href {\doibase 10.1073/pnas.1720291115}
  {\bibfield  {journal} {\bibinfo  {journal} {Proceedings of the National
  Academy of Sciences}\ }\textbf {\bibinfo {volume} {115}},\ \bibinfo {pages}
  {5343} (\bibinfo {year} {2018})},\ \Eprint
  {http://arxiv.org/abs/http://www.pnas.org/content/115/21/5343.full.pdf}
  {http://www.pnas.org/content/115/21/5343.full.pdf} \BibitemShut {NoStop}%
\bibitem [{\citenamefont {Steglich}\ \emph {et~al.}(2012)\citenamefont
  {Steglich}, \citenamefont {Arndt}, \citenamefont {Stockert}, \citenamefont
  {Friedemann}, \citenamefont {Brando}, \citenamefont {Klingner}, \citenamefont
  {Krellner}, \citenamefont {Geibel}, \citenamefont {Wirth}, \citenamefont
  {Kirchner},\ and\ \citenamefont {Si}}]{steglich_12}%
  \BibitemOpen
  \bibfield  {author} {\bibinfo {author} {\bibfnamefont {F.}~\bibnamefont
  {Steglich}}, \bibinfo {author} {\bibfnamefont {J.}~\bibnamefont {Arndt}},
  \bibinfo {author} {\bibfnamefont {O.}~\bibnamefont {Stockert}}, \bibinfo
  {author} {\bibfnamefont {S.}~\bibnamefont {Friedemann}}, \bibinfo {author}
  {\bibfnamefont {M.}~\bibnamefont {Brando}}, \bibinfo {author} {\bibfnamefont
  {C.}~\bibnamefont {Klingner}}, \bibinfo {author} {\bibfnamefont
  {C.}~\bibnamefont {Krellner}}, \bibinfo {author} {\bibfnamefont
  {C.}~\bibnamefont {Geibel}}, \bibinfo {author} {\bibfnamefont
  {S.}~\bibnamefont {Wirth}}, \bibinfo {author} {\bibfnamefont
  {S.}~\bibnamefont {Kirchner}}, \ and\ \bibinfo {author} {\bibfnamefont
  {Q.}~\bibnamefont {Si}},\ }\href@noop {} {\bibfield  {journal} {\bibinfo
  {journal} {Journal of physics: Condensed Matter}\ }\textbf {\bibinfo {volume}
  {24}} (\bibinfo {year} {2012})}\BibitemShut {NoStop}%
\bibitem [{\citenamefont {Bruls}\ \emph {et~al.}(1994)\citenamefont {Bruls},
  \citenamefont {Wolf}, \citenamefont {Finsterbusch}, \citenamefont
  {Thalmeier}, \citenamefont {Kouroudis}, \citenamefont {Sun}, \citenamefont
  {Assmus}, \citenamefont {L\"uthi}, \citenamefont {Lang}, \citenamefont
  {Gloos}, \citenamefont {Steglich},\ and\ \citenamefont {Modler}}]{bruls_94}%
  \BibitemOpen
  \bibfield  {author} {\bibinfo {author} {\bibfnamefont {G.}~\bibnamefont
  {Bruls}}, \bibinfo {author} {\bibfnamefont {B.}~\bibnamefont {Wolf}},
  \bibinfo {author} {\bibfnamefont {D.}~\bibnamefont {Finsterbusch}}, \bibinfo
  {author} {\bibfnamefont {P.}~\bibnamefont {Thalmeier}}, \bibinfo {author}
  {\bibfnamefont {I.}~\bibnamefont {Kouroudis}}, \bibinfo {author}
  {\bibfnamefont {W.}~\bibnamefont {Sun}}, \bibinfo {author} {\bibfnamefont
  {W.}~\bibnamefont {Assmus}}, \bibinfo {author} {\bibfnamefont
  {B.}~\bibnamefont {L\"uthi}}, \bibinfo {author} {\bibfnamefont
  {M.}~\bibnamefont {Lang}}, \bibinfo {author} {\bibfnamefont {K.}~\bibnamefont
  {Gloos}}, \bibinfo {author} {\bibfnamefont {F.}~\bibnamefont {Steglich}}, \
  and\ \bibinfo {author} {\bibfnamefont {R.}~\bibnamefont {Modler}},\ }\href
  {\doibase 10.1103/PhysRevLett.72.1754} {\bibfield  {journal} {\bibinfo
  {journal} {Physical Review Letters}\ }\textbf {\bibinfo {volume} {72}},\
  \bibinfo {pages} {1754} (\bibinfo {year} {1994})}\BibitemShut {NoStop}%
\bibitem [{\citenamefont {Steglich}\ \emph {et~al.}(2000)\citenamefont
  {Steglich}, \citenamefont {Gegenwart}, \citenamefont {Geibel}, \citenamefont
  {Hinze}, \citenamefont {Lang}, \citenamefont {Langhammer}, \citenamefont
  {Sparn},\ and\ \citenamefont {Trovarelli}}]{steglich_00}%
  \BibitemOpen
  \bibfield  {author} {\bibinfo {author} {\bibfnamefont {F.}~\bibnamefont
  {Steglich}}, \bibinfo {author} {\bibfnamefont {P.}~\bibnamefont {Gegenwart}},
  \bibinfo {author} {\bibfnamefont {C.}~\bibnamefont {Geibel}}, \bibinfo
  {author} {\bibfnamefont {P.}~\bibnamefont {Hinze}}, \bibinfo {author}
  {\bibfnamefont {M.}~\bibnamefont {Lang}}, \bibinfo {author} {\bibfnamefont
  {C.}~\bibnamefont {Langhammer}}, \bibinfo {author} {\bibfnamefont
  {G.}~\bibnamefont {Sparn}}, \ and\ \bibinfo {author} {\bibfnamefont
  {O.}~\bibnamefont {Trovarelli}},\ }\href {\doibase
  https://doi.org/10.1016/S0921-4526(99)01736-6} {\bibfield  {journal}
  {\bibinfo  {journal} {Physica B: Condensed Matter}\ }\textbf {\bibinfo
  {volume} {280}},\ \bibinfo {pages} {349 } (\bibinfo {year}
  {2000})}\BibitemShut {NoStop}%
\bibitem [{\citenamefont {Tayama}\ \emph {et~al.}(2003)\citenamefont {Tayama},
  \citenamefont {Lang}, \citenamefont {L\"uhmann}, \citenamefont {Steglich},\
  and\ \citenamefont {Assmus}}]{tayama_03}%
  \BibitemOpen
  \bibfield  {author} {\bibinfo {author} {\bibfnamefont {T.}~\bibnamefont
  {Tayama}}, \bibinfo {author} {\bibfnamefont {M.}~\bibnamefont {Lang}},
  \bibinfo {author} {\bibfnamefont {T.}~\bibnamefont {L\"uhmann}}, \bibinfo
  {author} {\bibfnamefont {F.}~\bibnamefont {Steglich}}, \ and\ \bibinfo
  {author} {\bibfnamefont {W.}~\bibnamefont {Assmus}},\ }\href {\doibase
  10.1103/PhysRevB.67.214504} {\bibfield  {journal} {\bibinfo  {journal}
  {Physical Review B}\ }\textbf {\bibinfo {volume} {67}},\ \bibinfo {pages}
  {214504} (\bibinfo {year} {2003})}\BibitemShut {NoStop}%
\bibitem [{\citenamefont {Lang}\ \emph {et~al.}(1999)\citenamefont {Lang},
  \citenamefont {Gegenwart}, \citenamefont {Helfrich}, \citenamefont
  {K\"oppen}, \citenamefont {Kromer}, \citenamefont {Langhammer}, \citenamefont
  {Geibel}, \citenamefont {Steglich}, \citenamefont {Kim},\ and\ \citenamefont
  {Stewart}}]{gonis_99}%
  \BibitemOpen
  \bibfield  {author} {\bibinfo {author} {\bibfnamefont {M.}~\bibnamefont
  {Lang}}, \bibinfo {author} {\bibfnamefont {P.}~\bibnamefont {Gegenwart}},
  \bibinfo {author} {\bibfnamefont {R.}~\bibnamefont {Helfrich}}, \bibinfo
  {author} {\bibfnamefont {M.}~\bibnamefont {K\"oppen}}, \bibinfo {author}
  {\bibfnamefont {F.}~\bibnamefont {Kromer}}, \bibinfo {author} {\bibfnamefont
  {C.}~\bibnamefont {Langhammer}}, \bibinfo {author} {\bibfnamefont
  {C.}~\bibnamefont {Geibel}}, \bibinfo {author} {\bibfnamefont
  {F.}~\bibnamefont {Steglich}}, \bibinfo {author} {\bibfnamefont {J.~S.}\
  \bibnamefont {Kim}}, \ and\ \bibinfo {author} {\bibfnamefont {G.~R.}\
  \bibnamefont {Stewart}},\ }\href {\doibase 10.1007/978-1-4615-4715-0} {\emph
  {\bibinfo {title} {Electron Correlations and Materials Properties}}},\
  \bibinfo {edition} {1st}\ ed.,\ edited by\ \bibinfo {editor} {\bibfnamefont
  {A.}~\bibnamefont {Gonis}}, \bibinfo {editor} {\bibfnamefont
  {N.}~\bibnamefont {Kioussis}}, \ and\ \bibinfo {editor} {\bibfnamefont
  {M.}~\bibnamefont {Ciftan}}\ (\bibinfo  {publisher} {Springer US},\ \bibinfo
  {year} {1999})\ pp.\ \bibinfo {pages} {153--168}\BibitemShut {NoStop}%
\bibitem [{\citenamefont {Thalmeier}\ and\ \citenamefont
  {Zwicknagl}()}]{thalmeier_05}%
  \BibitemOpen
  \bibfield  {author} {\bibinfo {author} {\bibfnamefont {P.}~\bibnamefont
  {Thalmeier}}\ and\ \bibinfo {author} {\bibfnamefont {G.}~\bibnamefont
  {Zwicknagl}},\ }\enquote {\bibinfo {title} {Unconventional superconductivity
  and magnetism in lanthanide and actinide intermetallic compounds},}\ in\
  \href@noop {} {\emph {\bibinfo {booktitle} {Handbook of the Physics and
  Chemistry of Rare Earth}}},\ Vol.~\bibinfo {volume} {34}\ (\bibinfo
  {publisher} {Elsevier})\ pp.\ \bibinfo {pages} {135--287}\BibitemShut
  {NoStop}%
\bibitem [{\citenamefont {Zwicknagl}(2016)}]{zwicknagl_16}%
  \BibitemOpen
  \bibfield  {author} {\bibinfo {author} {\bibfnamefont {G.}~\bibnamefont
  {Zwicknagl}},\ }\href {\doibase 10.1088/0034-4885/79/12/124501} {\bibfield
  {journal} {\bibinfo  {journal} {Reports on progress in physics}\ }\textbf
  {\bibinfo {volume} {79}},\ \bibinfo {pages} {124501} (\bibinfo {year}
  {2016})}\BibitemShut {NoStop}%
\bibitem [{\citenamefont {Sun}\ \emph {et~al.}(1990)\citenamefont {Sun},
  \citenamefont {Brand}, \citenamefont {Bruls},\ and\ \citenamefont
  {Assmus}}]{sun_90}%
  \BibitemOpen
  \bibfield  {author} {\bibinfo {author} {\bibfnamefont {W.}~\bibnamefont
  {Sun}}, \bibinfo {author} {\bibfnamefont {M.}~\bibnamefont {Brand}}, \bibinfo
  {author} {\bibfnamefont {G.}~\bibnamefont {Bruls}}, \ and\ \bibinfo {author}
  {\bibfnamefont {W.}~\bibnamefont {Assmus}},\ }\href@noop {} {\bibfield
  {journal} {\bibinfo  {journal} {Zeitschrift f{\"u}r Physik B Condensed
  Matter}\ }\textbf {\bibinfo {volume} {80}},\ \bibinfo {pages} {249} (\bibinfo
  {year} {1990})}\BibitemShut {NoStop}%
\bibitem [{\citenamefont {Stockert}\ \emph {et~al.}(2006)\citenamefont
  {Stockert}, \citenamefont {Andreica}, \citenamefont {Amato}, \citenamefont
  {Jeevan}, \citenamefont {Geibel},\ and\ \citenamefont
  {Steglich}}]{stockert_06b}%
  \BibitemOpen
  \bibfield  {author} {\bibinfo {author} {\bibfnamefont {O.}~\bibnamefont
  {Stockert}}, \bibinfo {author} {\bibfnamefont {D.}~\bibnamefont {Andreica}},
  \bibinfo {author} {\bibfnamefont {A.}~\bibnamefont {Amato}}, \bibinfo
  {author} {\bibfnamefont {H.~S.}\ \bibnamefont {Jeevan}}, \bibinfo {author}
  {\bibfnamefont {C.}~\bibnamefont {Geibel}}, \ and\ \bibinfo {author}
  {\bibfnamefont {F.}~\bibnamefont {Steglich}},\ }\href {\doibase
  10.1016/j.physb.2005.11.043} {\bibfield  {journal} {\bibinfo  {journal}
  {Physica B}\ }\textbf {\bibinfo {volume} {374-375}},\ \bibinfo {pages} {167}
  (\bibinfo {year} {2006})},\ \bibinfo {note} {proceedings of the Tenth
  International Conference on Muon Spin Rotation, Relaxation and
  Resonance}\BibitemShut {NoStop}%
\bibitem [{\citenamefont {Zhu}\ \emph {et~al.}(2003)\citenamefont {Zhu},
  \citenamefont {Garst}, \citenamefont {Rosch},\ and\ \citenamefont
  {Si}}]{zhu_03}%
  \BibitemOpen
  \bibfield  {author} {\bibinfo {author} {\bibfnamefont {L.}~\bibnamefont
  {Zhu}}, \bibinfo {author} {\bibfnamefont {M.}~\bibnamefont {Garst}}, \bibinfo
  {author} {\bibfnamefont {A.}~\bibnamefont {Rosch}}, \ and\ \bibinfo {author}
  {\bibfnamefont {Q.}~\bibnamefont {Si}},\ }\href@noop {} {\bibfield  {journal}
  {\bibinfo  {journal} {Physical Review Letters}\ }\textbf {\bibinfo {volume}
  {91}},\ \bibinfo {pages} {066404} (\bibinfo {year} {2003})}\BibitemShut
  {NoStop}%
\bibitem [{\citenamefont {Stockert}\ \emph {et~al.}(2004)\citenamefont
  {Stockert}, \citenamefont {Faulhaber}, \citenamefont {Zwicknagl},
  \citenamefont {St\"usser}, \citenamefont {Jeevan}, \citenamefont {Deppe},
  \citenamefont {Borth}, \citenamefont {K\"uchler}, \citenamefont
  {L\"owenhaupt}, \citenamefont {Geibel},\ and\ \citenamefont
  {Steglich}}]{stockert_04}%
  \BibitemOpen
  \bibfield  {author} {\bibinfo {author} {\bibfnamefont {O.}~\bibnamefont
  {Stockert}}, \bibinfo {author} {\bibfnamefont {E.}~\bibnamefont {Faulhaber}},
  \bibinfo {author} {\bibfnamefont {G.}~\bibnamefont {Zwicknagl}}, \bibinfo
  {author} {\bibfnamefont {N.}~\bibnamefont {St\"usser}}, \bibinfo {author}
  {\bibfnamefont {H.~S.}\ \bibnamefont {Jeevan}}, \bibinfo {author}
  {\bibfnamefont {M.}~\bibnamefont {Deppe}}, \bibinfo {author} {\bibfnamefont
  {R.}~\bibnamefont {Borth}}, \bibinfo {author} {\bibfnamefont
  {R.}~\bibnamefont {K\"uchler}}, \bibinfo {author} {\bibfnamefont
  {M.}~\bibnamefont {L\"owenhaupt}}, \bibinfo {author} {\bibfnamefont
  {C.}~\bibnamefont {Geibel}}, \ and\ \bibinfo {author} {\bibfnamefont
  {F.}~\bibnamefont {Steglich}},\ }\href@noop {} {\bibfield  {journal}
  {\bibinfo  {journal} {Physical Review Letters}\ }\textbf {\bibinfo {volume}
  {92}} (\bibinfo {year} {2004})}\BibitemShut {NoStop}%
\bibitem [{\citenamefont {Stockert}(2018)}]{stockert_18}%
  \BibitemOpen
  \bibfield  {author} {\bibinfo {author} {\bibfnamefont {O.}~\bibnamefont
  {Stockert}},\ }\href@noop {} {\enquote {\bibinfo {title} {private
  communication},}\ } (\bibinfo {year} {2018})\BibitemShut {NoStop}%
\bibitem [{\citenamefont {Zwicknagl}\ and\ \citenamefont
  {Pulst}(1993)}]{zwicknagl_93}%
  \BibitemOpen
  \bibfield  {author} {\bibinfo {author} {\bibfnamefont {G.}~\bibnamefont
  {Zwicknagl}}\ and\ \bibinfo {author} {\bibfnamefont {U.}~\bibnamefont
  {Pulst}},\ }\href {\doibase 10.1016/0921-4526(93)90736-P} {\bibfield
  {journal} {\bibinfo  {journal} {Physica B}\ }\textbf {\bibinfo {volume}
  {186}},\ \bibinfo {pages} {895} (\bibinfo {year} {1993})}\BibitemShut
  {NoStop}%
\bibitem [{\citenamefont {Hunt}\ \emph {et~al.}(1990)\citenamefont {Hunt},
  \citenamefont {Meeson}, \citenamefont {Probst}, \citenamefont {Reinders},
  \citenamefont {Springford}, \citenamefont {Assmus},\ and\ \citenamefont
  {Sun}}]{hunt_90}%
  \BibitemOpen
  \bibfield  {author} {\bibinfo {author} {\bibfnamefont {M.}~\bibnamefont
  {Hunt}}, \bibinfo {author} {\bibfnamefont {P.}~\bibnamefont {Meeson}},
  \bibinfo {author} {\bibfnamefont {P.-A.}\ \bibnamefont {Probst}}, \bibinfo
  {author} {\bibfnamefont {P.}~\bibnamefont {Reinders}}, \bibinfo {author}
  {\bibfnamefont {M.}~\bibnamefont {Springford}}, \bibinfo {author}
  {\bibfnamefont {W.}~\bibnamefont {Assmus}}, \ and\ \bibinfo {author}
  {\bibfnamefont {W.}~\bibnamefont {Sun}},\ }\href {\doibase
  10.1016/S0921-4526(90)81011-C} {\bibfield  {journal} {\bibinfo  {journal}
  {Journal of Physics: Condensed Matter}\ }\textbf {\bibinfo {volume} {2}},\
  \bibinfo {pages} {6859} (\bibinfo {year} {1990})}\BibitemShut {NoStop}%
\bibitem [{\citenamefont {Lifshitz}(1960)}]{lifshitz_60}%
  \BibitemOpen
  \bibfield  {author} {\bibinfo {author} {\bibfnamefont {I.~M.}\ \bibnamefont
  {Lifshitz}},\ }\href@noop {} {\bibfield  {journal} {\bibinfo  {journal}
  {Soviet Physics: Journal of Experimental and Theoretical Physics}\ }\textbf
  {\bibinfo {volume} {11}},\ \bibinfo {pages} {1130} (\bibinfo {year}
  {1960})}\BibitemShut {NoStop}%
\bibitem [{\citenamefont {K\"uchler}\ \emph {et~al.}(2003)\citenamefont
  {K\"uchler}, \citenamefont {Oeschler}, \citenamefont {Gegenwart},
  \citenamefont {Cichorek}, \citenamefont {Neumaier}, \citenamefont {Tegus},
  \citenamefont {Geibel}, \citenamefont {Mydosh}, \citenamefont {Steglich},
  \citenamefont {Zhu},\ and\ \citenamefont {Si}}]{kuechler_03}%
  \BibitemOpen
  \bibfield  {author} {\bibinfo {author} {\bibfnamefont {R.}~\bibnamefont
  {K\"uchler}}, \bibinfo {author} {\bibfnamefont {N.}~\bibnamefont {Oeschler}},
  \bibinfo {author} {\bibfnamefont {P.}~\bibnamefont {Gegenwart}}, \bibinfo
  {author} {\bibfnamefont {T.}~\bibnamefont {Cichorek}}, \bibinfo {author}
  {\bibfnamefont {K.}~\bibnamefont {Neumaier}}, \bibinfo {author}
  {\bibfnamefont {O.}~\bibnamefont {Tegus}}, \bibinfo {author} {\bibfnamefont
  {C.}~\bibnamefont {Geibel}}, \bibinfo {author} {\bibfnamefont {J.~A.}\
  \bibnamefont {Mydosh}}, \bibinfo {author} {\bibfnamefont {F.}~\bibnamefont
  {Steglich}}, \bibinfo {author} {\bibfnamefont {L.}~\bibnamefont {Zhu}}, \
  and\ \bibinfo {author} {\bibfnamefont {Q.}~\bibnamefont {Si}},\ }\href@noop
  {} {\bibfield  {journal} {\bibinfo  {journal} {Physical Review Letters}\
  }\textbf {\bibinfo {volume} {91}},\ \bibinfo {pages} {066405} (\bibinfo
  {year} {2003})}\BibitemShut {NoStop}%
\bibitem [{\citenamefont {K\"uchler}\ \emph
  {et~al.}(2004{\natexlab{a}})\citenamefont {K\"uchler}, \citenamefont
  {Gegenwart}, \citenamefont {Heuser}, \citenamefont {Scheidt}, \citenamefont
  {Stewart},\ and\ \citenamefont {Steglich}}]{kuechler_04a}%
  \BibitemOpen
  \bibfield  {author} {\bibinfo {author} {\bibfnamefont {R.}~\bibnamefont
  {K\"uchler}}, \bibinfo {author} {\bibfnamefont {P.}~\bibnamefont
  {Gegenwart}}, \bibinfo {author} {\bibfnamefont {K.}~\bibnamefont {Heuser}},
  \bibinfo {author} {\bibfnamefont {E.-W.}\ \bibnamefont {Scheidt}}, \bibinfo
  {author} {\bibfnamefont {G.~R.}\ \bibnamefont {Stewart}}, \ and\ \bibinfo
  {author} {\bibfnamefont {F.}~\bibnamefont {Steglich}},\ }\href@noop {}
  {\bibfield  {journal} {\bibinfo  {journal} {Physical Review Letters}\
  }\textbf {\bibinfo {volume} {93}},\ \bibinfo {pages} {096402} (\bibinfo
  {year} {2004}{\natexlab{a}})}\BibitemShut {NoStop}%
\bibitem [{\citenamefont {K\"uchler}\ \emph
  {et~al.}(2004{\natexlab{b}})\citenamefont {K\"uchler}, \citenamefont
  {Weickert}, \citenamefont {Gegenwart}, \citenamefont {Oeschler},
  \citenamefont {Ferstl}, \citenamefont {Geibel},\ and\ \citenamefont
  {Steglich}}]{kuechler_04b}%
  \BibitemOpen
  \bibfield  {author} {\bibinfo {author} {\bibfnamefont {R.}~\bibnamefont
  {K\"uchler}}, \bibinfo {author} {\bibfnamefont {F.}~\bibnamefont {Weickert}},
  \bibinfo {author} {\bibfnamefont {P.}~\bibnamefont {Gegenwart}}, \bibinfo
  {author} {\bibfnamefont {N.}~\bibnamefont {Oeschler}}, \bibinfo {author}
  {\bibfnamefont {J.}~\bibnamefont {Ferstl}}, \bibinfo {author} {\bibfnamefont
  {C.}~\bibnamefont {Geibel}}, \ and\ \bibinfo {author} {\bibfnamefont
  {F.}~\bibnamefont {Steglich}},\ }\href@noop {} {\bibfield  {journal}
  {\bibinfo  {journal} {Journal of Magnetism and Magnetic Materials}\ }\textbf
  {\bibinfo {volume} {272-276}},\ \bibinfo {pages} {229} (\bibinfo {year}
  {2004}{\natexlab{b}})},\ \bibinfo {note} {proceedings of the International
  Conference on Magnetism (ICM 2003)}\BibitemShut {NoStop}%
\end{thebibliography}%
\end{document}